\documentclass[11pt,english]{article}
\usepackage[T1]{fontenc}
\usepackage{color}
\usepackage{amsmath}
\usepackage{amssymb}
\usepackage{setspace}
\usepackage{esint}
\usepackage{multirow}
\makeatletter
\usepackage[utf8]{inputenc}
\usepackage{amsthm}
\usepackage{graphicx}
\usepackage{amsfonts}
\usepackage{tikz}
\usepackage{slashed}
\usepackage{braket}

\usepackage{afterpage}
\usepackage{cite}
\usetikzlibrary{calc}
\usepackage{color}

\usepackage{lipsum}
\usepackage{mathtools}
\usepackage{cuted}

\usetikzlibrary{decorations.markings}
\usepackage[margin=1in]{geometry}

\usepackage{babel}
\usepackage{multicol}

\makeatother

\usepackage{babel}

\newcommand{\abovedisplayline}{\vspace{-\multicolsep}\noindent
	\rlap{\rule[.6\baselineskip]{\dimexpr 0.5\textwidth-0.5\columnsep}{.5pt}}}

\newcommand{\belowdisplayline}{\hspace{\textwidth}%
	\llap{\rule{\dimexpr 0.5\textwidth-0.5\columnsep}{.5pt}}%
	\vspace{-0.2\baselineskip}\par\vspace{-\multicolsep}}

\makeatletter
\newenvironment{tablehere}
{\def\@captype{table}}
{}

\newenvironment{figurehere}
{\def\@captype{figure}}
{}
\makeatother

\begin{document}

	\title{Incompatible  coordinate algebra representations as the origin of particle generations}
	\author{S. Farnsworth}
	\maketitle
	 \begin{center}
   Max Planck Institute for Gravitational Physics (Albert Einstein Institute), Germany.
	\end{center} 

\begin{abstract}
	The success of the Higgs mechanism in the standard model has led to  the speculation that the  standard model gauge group might arise through an analogous breaking   of a yet more unified group. Such `grand unified theories' have the advantage of unifying both the gauge structure and fermion representations of the standard model. Unfortunately, the theories that most elegantly unify the fermions, without predicting extra unobserved fermion states, do not explain the existence of the three fermion generations. They also  typically predict a proliferation of bosonic states, which lead to so-far unobserved processes like proton decay. In this paper we introduce an alternative explanation for why one might only observe a subgroup of a larger `unified' group in nature. The approach we introduce   gives rise naturally to a generation structure without the appearance of unwanted fermion states, and is  cleaner  in the sense that it avoids the usual proliferation of unobserved bosonic states and resulting unobserved processes.
\end{abstract}

\begin{multicols}{2}
\section{Introduction}

The representations and charges observed in  the standard model fermion sector remain unexplained. Grand unified theories (GUTs) are perhaps the best known attempt in this direction, but   while such models present a conceptually satisfying explanation for the patterns and features of the standard model fermion sector,  they usually suffer from a proliferation of heavy bosonic fields, and often predict unobserved consequences such as  proton decay~\cite{Protdec}, and magnetic monopoles~\cite{Preskil}. In this paper we  point out that the fermion representations and generation structure of low energy physics are very analogous to what is observed in `incompatible' algebra representations (introduced below). Such representations are naturally incorporated into models that unify gravity with particle physics, by appropriately generalizing the coordinate algebra of Riemannian geometry to also simultaneously describe the Yang-Mills theories living on the spacetime \cite{Chamseddine:2007oz,farnsworth20}. We present several simple examples that, although they do not yet capture the full structure of the standard model and should be regarded as toy examples, nevertheless point to an intriguing new approach to breaking a GUT group down to the standard model gauge group, which has some advantages over the more familiar mechanism of breaking  via the  Higgs mechanism.

This paper is organized as follows. We start by defining incompatible representations, followed by which we provide a very simple example that displays symmetry breaking along with the appearance of `generations'. We then consider a slightly more realistic example that exhibits the group representations for three generations of standard model leptons. We finally conclude with a brief description of how this mechanism might arise in models of nature.



\section{Incompatible representations}

A representation $(\pi,H)$ of a (not necessarily associative) algebra $A$ is a map from elements $f\in A$ to  linear transformations $\pi(f)$ on a vector space $H$.  We define a representation to be  `compatible', if its corresponding `Eilenberg algebra' $B=A\oplus H$~\cite{Eilenberg,Schafer,FarBoy2015}, has the same associative properties as $A$. For example, a `left' representation $(\pi,H)$ of an associative algebra $A$ is compatible if it satisfies the associativity condition:
\begin{align}
[\pi(f),\pi(g),h] = 0, 
\end{align}
where $f,g\in A$, $h\in H$, and $[a,b,c] = (ab)c-a(bc)$ is the associator. Algebra representations are usually defined to be `compatible'. In this paper, however, we relax the usual definition, to consider representations $(\pi,H)$, for which the corresponding Eilenberg algebra $B=A\oplus H$, has different associative properties to the algebra $A$ that is being represented (although we still restrict to representations $\pi(f)$ that respect the identity on $A$, and the linearity of both $A$ and $H$). We call such representations `incompatible'.

The continuous symmetries of an algebra representation are generated by derivation maps $T:H\rightarrow H$, that satisfy:
\begin{align}
T\pi(f)h = \pi(\delta[f])h+\pi(f)Th, \label{derivaitons} 
\end{align}
where $f\in A$, and  $\delta\in Der(A)$ is a derivation of the algebra $A$~\cite{FarBoy2015}. Compatible representations typically preserve the symmetries of an algebra, such that there exists a generator $T$ satisfying Eq.~\eqref{derivaitons} for every derivation $\delta\in A$.  When considering incompatible representations, however, the mismatch between the associativity properties of an algebra and its representation can lead to a drastic reduction in symmetry, and can often lead to the development of `generation' structure. This is the mechanism that we exploit in this paper.

\section{A simple example}
	
In order to showcase the associative `mismatch' that we seek to describe,  we begin with a simple example based on the octonion algebra $A=\mathbb{O}$.  We introduce a representation that is incompatible with the `alternative' product of the octonions, resulting in a breaking of $G_2$ exceptional symmetry, down to an unbroken $SU_2$ symmetry, and the appearance of two `generations'.

A nice way of viewing the octonions, is through the Cayley-Dickson construction, which produces a  sequence of algebras, each of which is twice the dimension of the previous. The sequence begins with the real numbers $\mathbb{R}$, and then makes use of the real algebra to construct the complex numbers $\mathbb{C}$, and then the complex numbers to construct the quanternions $\mathbb{H}$, followed by the octonions $\mathbb{O}$, and so on. This process results in a well known $2\times 2$ matrix representation of both the complex numbers on $\mathbb{R}^2$, and the quaternions on $\mathbb{C}^2$:
\begin{subequations}
\begin{align}
c &= \begin{pmatrix}
a & b\\
-b & a
\end{pmatrix}, & a,b&\in\mathbb{R},\\
q &= \begin{pmatrix}
r & s\\
-\overline{s} & \overline{r}
\end{pmatrix}, & r,s&\in\mathbb{C},\label{QuaRep}
\end{align}
where `$c$' is a representation of a complex number,  `$q$' is a representation of a quanternion element, and where the bar ``$\overline{~}$'' represents complex conjugation. The complex and quaternion products correspond to  matrix multiplication, and so are naturally `compatible' with the matrix action on $\mathbb{R}^2$ and $\mathbb{C}^2$ respectively. 

One might hope to use the Cayley-Dickson construction to define a similar matrix representation of the octonions, with basis elements
\begin{align}
e_i &= \begin{pmatrix}
q_i & 0 \\
0 & q_i^*
\end{pmatrix}, & e_{i+4} & = \begin{pmatrix}
0& q_i  \\
-q_i^* & 0
\end{pmatrix},\label{OctRep}
\end{align}
\end{subequations}
where the $q_i\in\mathbb{H}$, $i=,0,1,2,3,$  form a basis for the quaternions, and where the $*$ denotes the usual involution on the quaternions.
Unfortunately, because the octonions are nonassociative,  the usual matrix product is inappropriate. We instead introduce a  `flipped' matrix product defined by:
\global\columnwidth=\columnwidth
\end{multicols}
\abovedisplayline
\vspace{-.35cm}
\begin{align}
\begin{pmatrix}
u_{11} & u_{12}\\
u_{21}&u_{22}
\end{pmatrix}
\begin{pmatrix}
v_{11} & v_{12}\\
v_{21}&v_{22}
\end{pmatrix}=
\begin{pmatrix}
u_{11}v_{11}+\color{blue}v_{21}u_{12}\color{black} & \color{blue}v_{12}u_{11}\color{black}+u_{12}v_{22}\\
\color{blue}v_{11}u_{21}\color{black}+u_{22}v_{21}&u_{21}v_{12}+\color{blue}v_{22}u_{22}\color{black}
\end{pmatrix},\label{prodrule}
\end{align}
\belowdisplayline

\begin{multicols}{2}

\noindent where each of the entries are quaternion elements (the quaternions are noncommutative, and so the ordering matters). When applied to matrix elements of the form given in Eq.~\eqref{OctRep}, an octonion product results, which is summarized by the Fano plane given in Fig.~\ref{Fano}. Furthermore, the octonion involution corresponds simply to the conjugate transpose. This representation therefore acts as an algebra $\star$-homomorphism, preserving the full structure of the octonions.

\begin{center}
	
	\begin{figurehere} 	
		\begin{tikzpicture}[scale=1.3,cap=round,>=latex]
		\draw[
		decoration={markings, mark = between positions .26 and 0.95 step 0.53 with {\arrow[very thick]{>}}},
		postaction={decorate}
		] (210:2)--(90:2)  ;
		\draw[
		decoration={markings, mark = between positions .26 and 0.95 step 0.53 with {\arrow[very thick]{>}}},
		postaction={decorate}
		] (90:2)  --(330:2);
		\draw[
		decoration={markings, mark = between positions .26 and 0.95 step 0.53 with {\arrow[very thick]{>}}},
		postaction={decorate}
		] (330:2)--(210:2);
		
		\draw[
		decoration={markings, mark = between positions .18 and 0.95 step 0.6 with {\arrow[very thick]{>}}},
		postaction={decorate}
		]  (30:1)  -- (210:2);
		\draw[
		decoration={markings, mark = between positions .18 and 0.95 step 0.6 with {\arrow[very thick]{>}}},
		postaction={decorate}
		]  (150:1) -- (330:2);
		\draw[
		decoration={markings, mark = between positions .18 and 0.95 step 0.6 with {\arrow[very thick]{>}}},
		postaction={decorate}
		]  (270:1) -- (90:2);
		
		\draw[
		decoration={markings, mark = between positions .32 and 0.99 step 0.33 with {\arrow[very thick]{<}}},
		postaction={decorate}
		]
		(0,0) circle (1);
		\filldraw[fill=white, draw=black] 
		(0:0)   circle(7pt)
		(30:1)  circle(7pt)
		(90:2)  circle(7pt)
		(150:1) circle(7pt)
		(210:2) circle(7pt)
		(270:1) circle(7pt)
		(330:2) circle(7pt);
		\node[circle] at (0:0) {$e_4$};
		\node[circle] at (30:1) {$e_2$};
		\node[circle] at (90:2) {$e_7$};
		\node[circle] at (150:1) {$e_1$};
		\node[circle] at (210:2) {$e_6$};
		\node[circle] at (270:1) {$e_3$};
		\node[circle] at (330:2) {$e_5$};
		\end{tikzpicture}
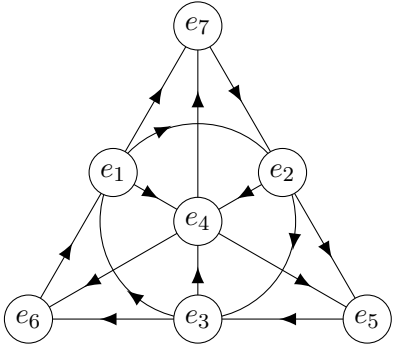
		\caption{The octonion product.}\label{Fano}
	\end{figurehere}
\end{center}

Represented as `matrices' as in Eq.~\eqref{OctRep}, the octonions can be equipped with a   matrix action on $\mathbb{C}^4$ (via the representation of the quaternions on $\mathbb{C}^2$ given in Eq.~\eqref{QuaRep}). This action is incompatible with the `flipped' matrix  product, and as such an immediate question is how much of the  symmetry of the product it preserves. This can be determined  by solving Eq.~\eqref{derivaitons} to find generators `$T$' corresponding to the elements of the derivation algebra of the octonions $\delta\in Der(\mathbb{O})$.

\noindent A convenient basis for $Der(\mathbb{O})$ is given by~\cite{Schaf49}:
\begin{align}
\delta_{ij} = [L_{e_i},L_{e_j}]+ [L_{e_i},R_{e_j}]+[R_{e_i},R_{e_j}]\label{deroct}
\end{align}
where $L_ab = ab$, and $R_ab = ba$ for $a,b\in\mathbb{O}$ are the standard `left-right' notations that are common place  when dealing with nonassociative algebras~\cite{Schafer}. The derivation algebra of the octonions is the $14$-dimensional exceptional Lie algebra $g_2$ (the basis given in Eq.~\ref{deroct} is degenerate)~\cite{Baez}. Not all derivation elements will   correspond to  solutions of Eq.~\eqref{derivaitons}, however. Instead, only an $su(2)$ sub-algebra will act as derivations on the  representation. A neat basis is provided by:
\begin{align}
\delta_1 &= \frac{2}{3}(\delta_{46}-2\delta_{75}),\nonumber\\
\delta_2 &= \frac{2}{3}(\delta_{47}-2\delta_{56}),\nonumber\\
\delta_3 &= \frac{2}{3}(\delta_{45}-2\delta_{67}).\nonumber
\end{align}
These basis elements corresponds respectively to following solutions of  Eq.~\eqref{derivaitons}:
\global\columnwidth=\columnwidth
\end{multicols}
\abovedisplayline
\begin{align}
\vspace{-.39cm}
T_1 &= \begin{pmatrix}
0&1 &0&0\\
-1&0&0&0\\
0&0&0&1\\
0&0&-1&0  
\end{pmatrix},&
T_2 &= \begin{pmatrix}
i&0 &0&0\\
0&-i&0&0\\
0&0&i&0\\
0&0&0&-i
\end{pmatrix}, &
T_3 &= \begin{pmatrix}
0&i &0&0\\
i&0&0&0\\
0&0&0&i\\
0&0&i&0  
\end{pmatrix}.\nonumber
\end{align}
\begin{multicols}{2}
These solutions can be found by first searching for the most general operator $T$, which leaves the representation given in Eq.~\eqref{OctRep} closed under commutation, before restricting to those maps which satisfy the Leibniz rule on the octonion algebra.
	
Notice  that two things have occurred here. First, while the `flipped' matrix representation of the octonions acts as an algebra homomorphism, the `regular' matrix action on $\mathbb{C}^4$ has broken the $g_2$ generating algebra  of the octonions down to an $su_2$ subalgebra (i.e. the choice of representation picks out an  $su_2$ subalgebra). The representation space has furthermore split  into two distinct `generations' (i.e. two copies of the complex 2-dimensional representation of $SU_2$). 
\section{Chiral Models}

We next  consider a slightly more realistic scenario based on the   exceptional Jordan algebra of $3\times 3$, Hermitian, octonionic matrices $A=H_3(\mathbb{O})$. This algebra has $F_4$  exceptional symmetry, and is particularly interesting for our purposes because its  only  compatible representation is when it acts in the obvious way on itself~\cite{Jacob}. It has  also  caught a lot of  recent attention because of the way the standard model gauge group and the standard model fermion representations are  naturally embedded within it~\cite{Boyle20,Dubois20}. 
In this section we show how to  construct an incompatible representation of the exceptional Jordan algebra, which captures the symmetries and hypercharges of  three generations of standard model leptons.

A general element $\omega\in H_3(\mathbb{O})$ can be expressed as
\begin{align}
\omega = \begin{pmatrix}
a & C & B^\star\\
C^\star & b & A\\
B & A^\star&c
\end{pmatrix},
\end{align}
where $a,b,c\in\mathbb{R}$, and $A,B,C\in \mathbb{O}$, and where `$\star$' is the involution on the octonions. This $3\times 3$ construction implies a rather natural `incompatible' matrix representation on $H=\mathbb{O}^3$, which makes   use of the octonion product `element-wise'.  The continuous  symmetries of this representation are determined by solving Eq.~\eqref{derivaitons} to find the generating algebra $Der(A\oplus H)$.  Just as in the previous octonion example, not all derivations $\delta\in Der(H_3(\mathbb{O}))$, will correspond to solutions `$T$' of  Eq.~\eqref{derivaitons}. Instead,  only an $su_2\oplus g_2\le Der(H_3(\mathbb{O}))$ subalgebra survives, with the $24$ dimensional, real vector  space $H$ breaking up as $(3,7)\oplus(3,1)$.

 The $su_2\oplus g_2$ generating algebra on $A$, is expressible in the following convenient basis:
\begin{align}
\delta_{ij} &= \sum_{a=1,2,3} [L_{E_i^{(a)}},L_{E_j^{(a)}}],\label{extra0}\\
	\delta_{a} &=  [L_{E_8^{(a)}},L_{E_0^{(a)}}],\label{extra} 
	\end{align}
where the $E_j^{(a)}\in H_3(\mathbb{O})$, are basis elements in each of the three subalgebras $H^a_2(\mathbb{O}) = (\mathbb{I}- D_{(a)})H_3(\mathbb{O})(\mathbb{I}- D_{(a)})$, where $D_{(a)}$ is a mostly all zero $3\times 3$ matrix, with a `$1$' at position `$a=1,2,3$' along the diagonal. Each $E_j^{(a)}\in H_3(\mathbb{O})$ can be expressed compactly as:
\begin{align}
E_8^{(\bullet)} &= \begin{pmatrix}
1&0\\
0&-1
\end{pmatrix},& E_i^{(\bullet)} &= \begin{pmatrix}
0&e_i\\
e_i^*&0
\end{pmatrix},\label{embed} 
\end{align}
for $e_i\in\mathbb{O}$, $i=0,...,7$ (although it should not be forgotten that each $E_j^{(a)}$ is an element of $H_3(\mathbb{O})$). Just as in Eq.~\eqref{deroct}, the basis given in Eq.~\eqref{extra0} is degenerate, with only $14$ of the  $21$ possible elements being linearly independent. The three basis elements given in Eq.~\eqref{extra} correspond to the $su_2$ factor in the generating algebra.

Notice that the above group theory is readily distinguishable from what  would occur in a grand unified theory based on the $F_4$ Lie group. Under a breaking of the $f_4$ generating algebra down to  $su_2\oplus g_2$, a $26$ dimensional fermion representation would split into  $(3,7)\oplus (5,1)$. Furthermore, the $52$ gauge generators would split into the $17$ unbroken generators of $SU_2\times G_2$, together with $35$ broken generators corresponding to heavy bosonic states. In contrast, the above `incompatible' representation on $H=\mathbb{O}^3$  results in `fermion' states sitting in a $(3,7)\oplus (3,1)$ representation of the $su_2\oplus g_2$ generating algebra, while the $17$ gauge generators appear alone, without  the presence of the $35$ `broken' generators.

In the above example, the subalgebra $ su_2\oplus g_2\le Der(H_3(\mathbb{O}))$ has been preserved by making use of an action on $H=\mathbb{O}^3$, which preserves the structure of the octonion product  `element-wise'. The  $g_2$ factor, however, can be broken further to an $ su_2\oplus u_1$ subalgebra, under which a $7+1$ dimensional representation breaks into $(1)_0\oplus(1)_2\oplus(2)_{-1}$,  matching the chiral representations under which a single generation of leptons transform (including a right handed neutrino). It is natural to wonder if this more complete symmetry breaking might be achieved by further augmenting the action on $H$. To achieve this goal, we  introduce a matrix representation on $H$, with the element-wise  `bison' product given in table~\ref{Tab1}, replacing the previous element-wise octonion product. Bison algebras `$B_{(i)}$', $i=1,2,3,4,$ are $8$-dimensional, real, Division algebras very similar to the octonions, but which have $SU_2\times U_1$ symmetry~\cite{Toro,Bison} (the reader should take note that the labelling of the basis elements given in Table~\ref{Tab1} differs slightly from that given in~\cite{Toro}, and has been chosen to preserve as much symmetry as possible when matched to the octonion product given in Figure~\ref{Fano}).


\setlength{\tabcolsep}{4.pt}
\begin{tablehere}	
	
	\begin{center}
	\begin{tabular}{ |c||c|c|c|c|c|c|c | c| } 
		\hline
$e_0$ & $e_0$ & $e_1$& $e_2$& $e_3$& $e_4$&$e_5$&$e_6$&   $e_7$\\
\hline
\hline
$e_0$ & $e_0$ & $e_1$& $e_2$& $e_3$& $e_4$& $e_5$&$e_6$&   $e_7$\\
$e_1$ & $e_1$ & $e_2$& $e_3$& $-e_0$& $e_7$&   $-e_4$&$e_5$& $e_6$\\
$e_2$& $e_2$ & $-e_3$& $-e_0$& $e_1$&  $e_6$& $e_7$&$-e_4$&  $-e_5$\\
$e_3$ & $e_3$ & $e_0$& $-e_1$& $e_2$& $-e_5$&$-e_6$&  $e_7$& $-e_4$\\
$e_4$&$ e_4$ &$ e_7$&$ -e_6$&$ -e_5$&$ -e_0$& $ e_3$&$ e_2$& $-e_1$\\
$e_5$& $e_5$ & $e_4$&$ -e_7$&$ e_6$&$ e_3$&$ e_2$&$ -e_1$&$e_0$\\
$e_6$ & $e_6$ & $e_5$& $e_4$& $e_7$&  $-e_2$&$-e_1$&$-e_0$&  $-e_3$\\
$e_7$&$ e_7$ &$ -e_6$&$ e_5$&$ e_4$&$ -e_1$&$ -e_0$&$ -e_3$&$ e_2$\\
\hline 
	\end{tabular}
\caption{The second Bison action~\cite{Toro}.}
\label{Tab1}
\end{center}
\end{tablehere}
Under this new `incompatible' representation, which we denote $H=B_{(2)}^3$, only an $su_2\oplus su_2\oplus u_1$  derivation algebra survives on the corresponding Eilenberg algebra $B=A\oplus H$, with the representation space $H$ breaking up into $(3,1)_0\oplus(3,1)_2\oplus(3,2)_{-1}$. This representation corresponds to that of three `generations' of standard model leptons  (including right handed neutrinos), each related  by an additional $SU_2$ symmetry. The surviving `electro-weak' generators, can be expressed in  terms of the elements in Eq.~\eqref{extra0}, as:
\begin{align}
\delta_0 &= 2\delta_{46},\\
\delta_1 &= \frac{2}{3}(\delta_{26}+2\delta_{37}),\\
\delta_2 &= \frac{2}{3}(\delta_{24}+2\delta_{53}),\\
\delta_3 &= \frac{2}{3}(\delta_{46}+2\delta_{57}),
\end{align}
while the generators given in Eq.~\eqref{extra} remain unbroken. The additional, unobserved, $SU_2$ symmetry can also be broken, although the required representation is slightly more complicated to describe, and involves enlarging $H$ to include `anti-particle' states (by making direct use the three $H_2(\mathbb{O})$ embeddings given in Eq.~\eqref{embed}, each of which is isomorphic to the $9$-dimensional spin factor algebra).

\section{Physical interpretation}
 
The standard model fermion content is collected together into three `generations' that break up under the standard model gauge group into left and right handed quarks and leptons, each of which has unique hypercharge assignments. One possibility is that these observed patterns and features arise  dynamically, via the Higgs mechanism, from the representations of some larger group. An alternative approach, however, is to take the fermion content of the standard model at face value. 
If we view the standard model representations as fundamental, then a key  possibility is that they describe (and are unified as) an `internal' topological space that is characterized by an incompatible coordinate algebra representation (just as the `external' geometry is described by a Riemannian manifold characterized by the representation of a smooth coordinate algebra~\cite{Connes:2008vs}).

The idea that the universe might host such an incompatibility is  actually not so unfamiliar. When introducing particle fields into a physical theory, they should transform as representations of the symmetries of the theory. This leads to an interesting problem when including  spinors in theories of gravity because there are no spinor representations of manifold  diffeomorphisms. The solution is to introduce additional structure, namely a local frame with respect to which  spinors transform  under local Lorentz. When attempting to coordinatize the internal sector of a gauge theory (as is done for example in~\cite{Chamseddine:2007oz,farnsworth20}), it is natural to wonder what the analogous solution  might  be if particle content is introduced that does not sit in any `compatible' representation of the internal coordinatization (for example, if the universe wishes to describe `spinors' as a module over the exceptional Jordan algebra, which does not have `spinor' representations). 
One possibility, of course, is to form a tensor product with a representation that is compatible. 
The alternative possibility proposed here, however, is that  an `incompatible' representation could result, which preserves only some of the symmetry of the coordinatizing algebra, and splits the particle content into chiralities, leptons, quarks, and generations.\\

We would like to thank Latham Boyle,  Hadi Godazgar, and 
  Axel Kleinschmidt for helpful discussions. Special thanks  goes to Fredy  Jimenez who first introduced us to Bison algebras, and who has also begun exploring their application in particle model building~\cite{Bison}. This research was supported by the Max-Planck Institute.

\end{multicols}

\end{document}